\documentclass[amsmath,amssymb,nofootinbib,onecolumn,floatfix,12pt]{revtex4-1}
\usepackage{graphicx}   
\usepackage{verbatim}   
\usepackage{bm}
\usepackage{dcolumn}
\usepackage{textcomp}
\usepackage{rotating}
\usepackage{array}
\usepackage{accents}

\newcommand{\reaction}[2]{\overset{#1}{\underset{#2}{\rightleftharpoons}}}

\begin{document}
\title{Studying Protein Assembly with Reversible Brownian Dynamics of Patchy Particles}

\author{Heinrich C. R. Klein}
\affiliation{Institute for Theoretical Physics, Heidelberg University, 69120 Heidelberg, Germany}
\author{Ulrich S. Schwarz}
\email{ulrich.schwarz@bioquant.uni-heidelberg.de}
\affiliation{Institute for Theoretical Physics, Heidelberg University, 69120 Heidelberg, Germany}
\affiliation{BioQuant, Heidelberg University, 69120 Heidelberg, Germany}

\date{\today}

\begin{abstract}
Assembly of protein complexes like virus shells, the centriole, the
nuclear pore complex or the actin cytoskeleton is strongly
determined by their spatial structure. Moreover it is becoming increasingly
clear that the reversible nature of protein assembly is also an
essential element for their biological function. Here we introduce a
computational approach for the Brownian dynamics of patchy particles with
anisotropic assemblies and fully reversible reactions. Different particles
stochastically associate and dissociate with microscopic reaction rates
depending on their relative spatial positions. The translational and rotational
diffusive properties of all protein complexes are evaluated on-the-fly. Because
we focus on reversible assembly, we introduce a scheme which ensures
detailed balance for patchy particles. We then show how the macroscopic rates
follow from the microscopic ones. As an instructive example, we study the
assembly of a pentameric ring structure, for which we find excellent
agreement between simulation results and a macroscopic 
kinetic description without any adjustable parameters. This demonstrates
that our approach correctly accounts for both the diffusive
and reactive processes involved in protein assembly.
\end{abstract}


\maketitle

\section{Introduction}

Assembly of biomolecules into supramolecular complexes is at the heart
of many biological processes and the dynamic interplay of the
different components leads to biological functionality. The most
important type of self-assembling biomolecules are proteins. Although
they often have a globular shape, protein assemblies can be strongly
anisotropic due to the localized binding interactions between the
proteins. Like biological systems in general, protein assemblies are
operative on many different scales. Their size ranges from the
nanometer-scale (for example for two-component complexes like Barnase
and Barstar \cite{Gabdoulline1997}) through tens of nanometers (for
example for viruses, which typically consist of small multiples of 60
components \cite{Roos2010,Zlotnick2011}) up to the micrometer scale
(for example the mitotic spindle \cite{Compton2000,Karsenti2006,Gonczy2012}).
Even in steady state most biological complexes remain highly dynamic,
with association events being balanced by dissociation
events. Prominent examples are nuclear pore complexes
\cite{Alber2007,Angelo2008,Tran2006} or focal
adhesions\cite{Geiger2001,Bershadsky2006,Wolfenson2009}.
Another important example for the dynamic nature
of biological complexes are actin filaments
\cite{Beltzner2008,Guo2009,Pollard2007}, which are called
\textit{living polymers} due to their continuous exchange dynamics.
\cite{Guo2010} Similar to the broad range of length scales,
association rate constants observed in biological systems span a wide spectrum
ranging from $<10^3 \text{ M}^{-1}\text{s}^{-1}$ up
to $>10^{9} \text{ M}^{-1}\text{ s}^{-1}$. \cite{Schreiber2009}
The highly dynamic nature of biological assemblies as well as the
large range of involved spatial and temporal scales renders this
physical problem challenging but fascinating.

Interestingly, recent advances in the fabrication of functionalized
colloids with directional interactions (\textit{patchy particles})
make it possible to design elementary building blocks of micrometer
sizes which can be used to self-assemble complexes with new
functionality. Experimental techniques ranging from DNA-mediated
self-assembly \cite{Mirkin1996,Michele2013a,Wang2012,Knorowski2011} to
entropic depletion interactions \cite{Sacanna2010,Sacanna2013} have
been rapidly advancing during the last decade, thus providing a
plethora of possibilities to fabricate colloidal particles whose shape
and interactions can be controlled in detail. Moreover external
stimuli such as temperature, light or pH can be used to control the
inter-particle interactions during the assembly process
. \cite{Leunissen2009,Michele2013} These techniques allow for a
state-
or time-dependent switching of the interactions and can be used to
steer the assembly process. Controlling the particle
interactions during the assembly process can prevent kinetic trapping
resulting in a higher yield of the desired
structure, as has recently been shown in a computational study
for virus assembly. \cite{Baschek2012} To fully exploit the potential
of these techniques, a detailed understanding of the dynamics of the
assembly process (distribution of intermediates, relevant time scales)
is of crucial importance. 

Understanding the mechanisms governing chemical reactions has a long
tradition in theoretical physics and chemistry. The most powerful
analytical technique in this context is the Fokker-Planck or
Smoluchowski equation, which has been used early to study bimolecular
association based on diffusive motion.  This approach was pioneered by
Smoluchowski who first calculated the maximum diffusion-limited
reaction rate for a fully reactive spherical particle
. \cite{Smoluchowski1917} An important generalization of this result
was derived by Collins and Kimball \cite{Collins1949} who studied the
effect of finite reactivity by introducing a radiation boundary
condition relating the concentration at contact to the reactive
flux. Another essential extension is the case of particles with
anisotropic reactivity (\textit{patchy particles})
. \cite{Solc1973,Shoup1981,Berg1985,Shushin1999,Schlosshauer2002,Schlosshauer2004}
However, most of these generalizations focus on the calculation of
association rates and only a few aim at describing reversible
reactions. \cite{Agmon1984,Agmon1990} 

To study the full dynamics of protein assembly one needs to consider both association and
dissociation processes. Moreover, even for globular proteins the
intermediates formed during the assembly process are often of highly
non-spherical geometry, thus rendering an analytical treatment of
assembly in the framework of the Fokker-Planck equation very
difficult. In this case computer simulations provide a valuable
alternative.  While detailed molecular dynamics (MD) simulations have been
successfully used to investigate the behavior of a single molecule in great
detail, studying the dynamics of large protein complexes 
on biologically relevant time and length scales
is prohibited by the high computational costs of these simulations. Therefore
coarse-grained models are required for this case. Brownian dynamics (BD)
simulations, the numerical counterpart to the Fokker-Planck equation, have been
extensively used to study bimolecular association kinetics with
realistic protein
shapes. \cite{Northrup1984,Northrup1992,Zhou1990,Zhou1993,Gabdoulline2001,Gabdoulline1997,Zou2003,Alsallaq2007,Alsallaq2007a,Qin2011}
These studies have been focused mainly on a detailed calculation of
association rates in bimolecular reactions based on realistic protein
shapes and binding interactions. To study reaction dynamics in a large
system consisting of many proteins, various simulation frameworks have
been developed. \cite{Klann2012} They range from space- and time-continuous BD
simulations as for example in the Smoldyn framework
\cite{Andrews2004,Andrews2005} through event-driven tools like Greens
functions reaction dynamics (GFRD)
\cite{vanZon2005,Morelli2008,Takahashi2010}, which is based on the
analytical solution of the Fokker-Planck equation, up to a
space-discretized version of Gillespie's chemical master equation
approach \cite{Gillespie1977}, as for example used in MesoRD
\cite{Hattne2005,Fange2010,Fange2012}. While these simulation
frameworks have been successfully used to study reaction kinetics on a
large scale, they cannot be used to study the details of assembly
processes as all of these frameworks lack a detailed description of
protein anisotropy, including their shape or directional
interactions. To study the assembly of virus capsids as a paradigm of
a biological self-assembly process for which these elements are
essential, coarse-grained MD simulations
\cite{Rapaport2008,Hagan2006,Nguyen2007,Rapaport2012} and Monte Carlo (MC)
techniques \cite{Johnston2010,Horejs2011,Wilber2007,Wilber2009} have
been used. Similarly the interaction of colloidal particles has been
investigated on various scales with different simulation techniques
\cite{Knorowski2011}, including MD simulations
\cite{Largo2007},
MC simulations \cite{Damasceno2012,Liu2013} and BD studies
\cite{Sciortino2008,Halverson2013}.

Here we introduce a simulation framework which allows us to study the
spatial and stochastic aspects of protein assembly to large detail and
nevertheless is computationally relatively cheap. In our approach, we
combine BD of realistic protein shapes with stochastic
reactivity for both, association and dissociation. Proteins
are described as assemblies of spheres equipped with reactive
patches. Their motion in real and orientation space is based on the
overdamped Langevin equation with an anisotropic diffusion
tensor. Inspired by the notion of an encounter complex
\cite{Collins1949,Eigen1974,Shoup1982,Agmon1990,Gabdoulline1997,Alsallaq2007,Schreiber2009},
we decompose the reaction process into a diffusion and a reaction
part. Upon the diffusive formation of an encounter, two particles can
stochastically react with a microscopic reaction rate and thus form a
bond. Similarly an existing bond can be disrupted with a microscopic
dissociation rate. We first show how these microscopic reaction rates
can be related to macroscopic, experimentally measurable rates. We
then verify that our algorithm correctly reproduces the
macroscopically expected equilibrium behavior for bimolecular
reactions.  Finally we investigate the assembly of a pentameric ring
structure and compare our simulation results to a macroscopic rate
equation approach. Here we again find excellent agreement between the
stochastic simulations and the macroscopic description if the
macroscopic rates are calculated without any free parameter in the
correct way that includes both the diffusive and reactive
contributions. These results show the importance of including spatial
and orientational effects as well as realistic diffusion properties to
understand the dynamics governing protein assembly.

\section{Model and Methods}
\begin{figure}
    \includegraphics[width=8.5cm]{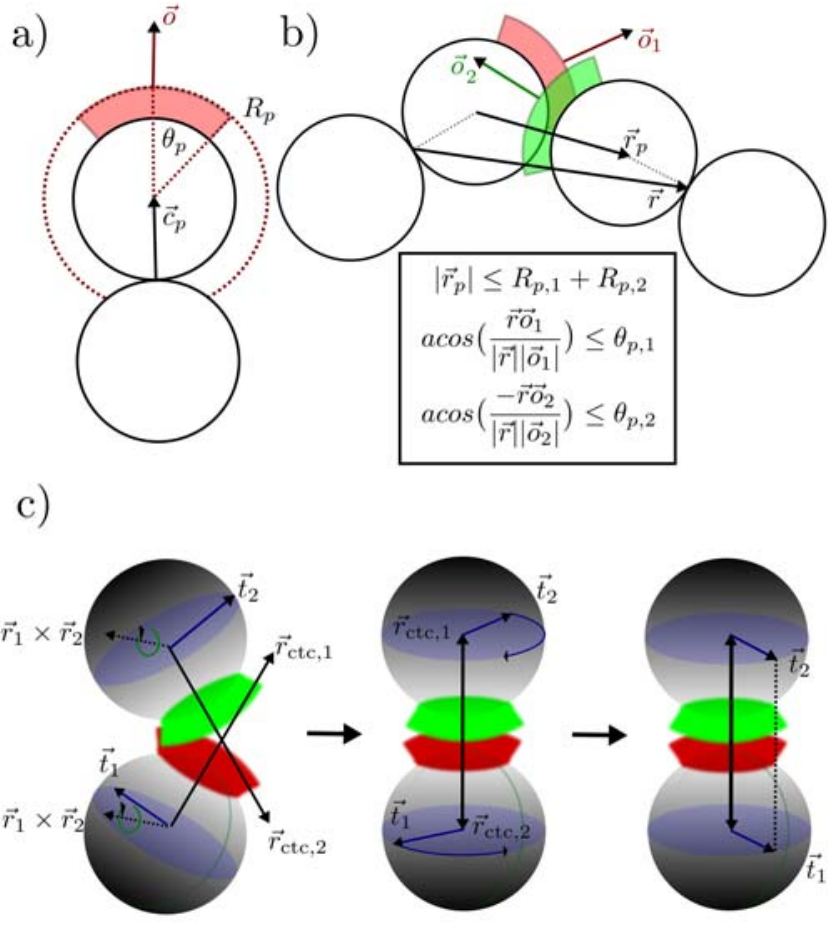}
    \caption{\label{Fig.1} Model definition. a) A dumbbell-shaped protein modeled by
two hard spheres. The protein is covered with a patch of radius $R_p$
whose center coincides with the center of one of the steric spheres
and is thus shifted by $\vec{c}_p$ relative to the center of the
protein (contact point of the spheres). The patch has an opening angle
of $\theta_p$ around the orientation vector $\vec{o}$ pointing along
the long axis of the dumbbell. b) The encounter between two of the
proteins of a). Here $\vec{r}$ is the
center-to-center(ctc)-vector between the proteins and $\vec{r}_p$ is
the center-to-center vector between the patches. c) Illustration of the reorientation
processes upon reaction. In a first step the ctc vectors of the
proteins are aligned with the desired ctc-vectors.  In a next step the
projection of the torsion vectors $\vec{t}_1$ and $\vec{t}_2$ on a
plane perpendicular to $\vec{r}_{\text{ctc}}$ are aligned. Finally the
clusters are shifted so that the desired relative distance between the
proteins is achieved. }
\end{figure}

\subsection{Encounter complex}

Our simulation approach is based on the concept of the
encounter complex. \cite{Collins1949,Eigen1974,Shoup1982,Agmon1990,Gabdoulline1997,Alsallaq2007,Schreiber2009}
In this concept a bimolecular reaction is decomposed into two steps:
the undirected diffusive motion of the two binding partners $A$ and
$B$ until they stochastically reach the encounter state $A\cdot B$ and
the reaction from the encounter state to the bound complex $C$. In
terms of the free energy landscape the encounter complex represents a
barrier separating the flat diffusive energy landscape from the
reaction funnel. Thus, it is a transient state which can be
characterized by the onset of highly correlated motion between the
binding partners. \cite{Alsallaq2007} A barrier in the free energy
landscape might arise for example due to a necessary rearrangement of
the binding site or due to a reorganization of the water layer. In
some cases there might not exist an explicit barrier, but even then
the encounter complex is a helpful theoretical concept. Considering
the encounter complex as the ``watershed'' between 
diffusion and reaction processes, the bimolecular reaction can be
split according to the following reaction scheme:
 \begin{align}
   A+B&\reaction{k_{+}}{k_{-}} C \\
   A+B&\reaction{k_{D}}{k_{D,b}} A\cdot B \reaction{k_a}{k_d} C \text{.}
\label{Eq. encounter} 
\end{align}
Here $k_+$ and $k_-$ represent the overall kinetic rates of the
reaction. In Eq.~\ref{Eq. encounter} the binding and unbinding process
is split into a diffusive part represented by the rates $k_D$ and
$k_{D,b}$ and a reaction part represented by $k_a$ and $k_d$,
respectively. Separating the binding process in free diffusive motion
without steering forces and localized binding requires short-ranged
interactions. Thus, our approach for the simulation of protein
association corresponds to a regime of high salt concentration
screening long-ranged electrostatic forces. Indeed this is a reasonable assumption 
for physiological salt conditions. Moreover, our approach can be used
to study the assembly of $\mu$m-sized colloids functionalized with
reactive patches where the size of the reactive region is small
compared to the particle size. As we consider a situation in which the
interaction between the proteins is short-ranged and only affects the
reaction probabilities in the encounter state, individual clusters
undergo free diffusive motion without an additional drift
term. Moreover, here we do not consider hydrodynamic interactions
which also might influence the binding kinetics.

\subsection{Patchy particles}

Our simulation system can be decomposed into four elementary
structures: proteins, patches, bonds, clusters.  Clusters are rigid
objects that can consist of a single or of multiple proteins held
together by bonds. Two clusters can react with each other by bond
formation between adhesive patches. A cluster consisting of multiple
proteins can decay into smaller clusters by bond
dissociation. Proteins are described by sets of non-overlapping, hard
spheres approximating their shape. The proteins are equipped with reactive
patches which reflect the localized binding sites. A patch is defined
starting from a sphere of radius $R_p$ whose center $\vec{c}_{p}$
(relative to the protein) can be chosen independent of the center of
the protein. In addition, each patch is described by an orientation
vector $\vec{o}$. As a simple example
for a non-globular protein, in Fig.~1a we schematically show a
dumbbell-shaped protein consisting of two hard spheres in contact. The
protein is equipped with one reactive patch of radius $R_p$ whose
center coincides with the center of one of the steric spheres and is
thus shifted by $\vec{c}_p$ from the protein center located at the
contact point of the steric spheres. The orientation vector $\vec{o}$
points along the long axis of the dumbbell. Using the above
definitions we formulate the following set of equations to define the
encounter between two clusters mediated by a pair of patches, compare
Fig.~1b:
\begin{align}
&r_p\leq R_{p,1}+R_{p,2} \label{Eq. Encounter1}\\
&\text{acos}\big(\frac{\vec{r}\cdot\vec{o}_1}{|\vec{r}||\vec{o}_1|}\big)\leq \theta_{p,1}  \text{, }
\text{acos}\big(\frac{-\vec{r}\cdot\vec{o}_2}{|\vec{r}||\vec{o}_2|}\big)\leq
\theta_{p,2} \text{.} \label{Eq. Encounter2} 
\end{align}
Eq.~\ref{Eq. Encounter1} means that the distance $r_p$ between the
centers of the two patches has to be sufficiently small for an
encounter to occur and can be implemented easily in a
simulation. Eq.~\ref{Eq. Encounter2} is more complex. It involves the
center-to-center vector $\vec{r}$ and not only reflects that the
patches are anisotropic, but also assumes that an encounter only
occurs if the two partners have a favorable orientation relatively to
each other which is defined by the two parameters $\theta_{p,1}$ and
$\theta_{p,1}$. In general an anisotropic protein can be
described by multiple spheres and the center of the patches can be
chosen independent of the steric spheres, allowing for a variable
description of proteins and their localized interactions.

\subsection{Particle motion}
\begin{figure}
    \includegraphics[width=8.5cm]{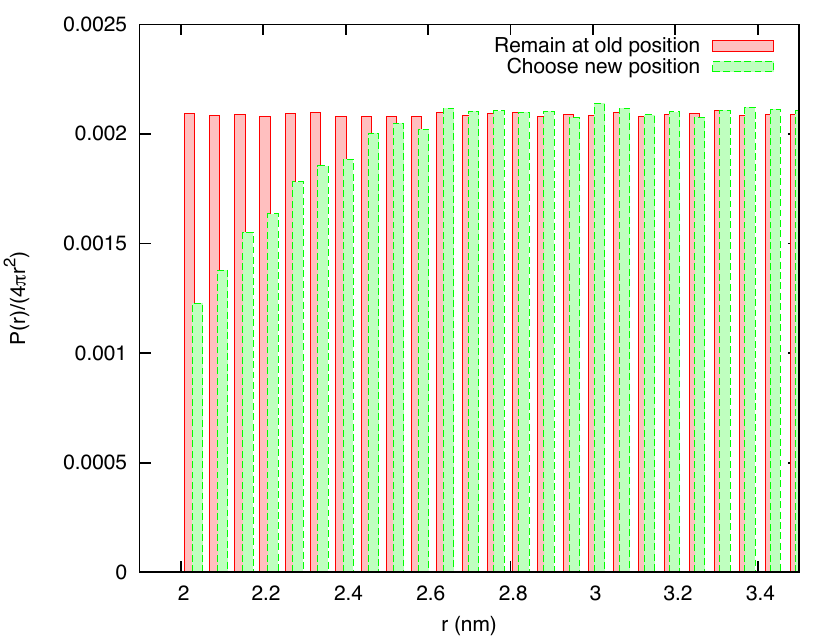}
     \caption{\label{Fig.2} Rejection methods. The distance distribution of two hard
spheres of radius $R=1$nm in a periodic box of volume
$V_\text{box}=(8\text{nm})^3$ is shown for two different rejection
methods. In the first rejection method (green) new positions
of the particles are chosen (starting from their positions before
collision) until no overlap is observed. In the second rejection
method (red) the spheres are set back to their position before the
collision. The simulations were performed at a time-step
resolution of $\Delta t=0.1$ns. }
\end{figure}

In our approach particle motion is described by the six-dimensional,
overdamped Langevin equation (Brownian motion) describing
translational and rotational diffusion of an arbitrarily shaped
but rigid object \cite{Korn2007,Schluttig2008}:
\begin{align}
&\partial_t  \mathbf{X}_t=\mathbf{g}_t \text{ with: } \langle
\mathbf{g}_t\rangle=0\text{, } \langle \mathbf{g}_t \mathbf{g}_{t^\prime}
\rangle=2k_B T \mathbf{M} \delta (t-t^\prime) \text{.}
\end{align}
Here $\mathbf{X}_t$ is a six-dimensional position vector describing the position
and orientations at time t and $\mathbf{g}_t$ is Gaussian 
white noise. $\mathbf{M}$ is the mobility matrix which is calculated on-the-fly for
any cluster shape \cite{Korn2007,Schluttig2008,Schluttig2010} following the
method of de la Torre \cite{GarciaDeLaTorre2000}.
Because intermediates continuously grow and shrink due
to association and dissociation, the diffusive
properties of the involved clusters are continuously changing
during a simulation. 

Protein particles are modeled as hard spheres and their collision
requires special attention. In a BD framework, particle
velocities are not defined and thus a ballistic reflection scheme is
not appropriate. Here we treat steric collisions by a rejection method similar to
MC simulations. If a steric overlap between two clusters is
created during a move step, this move step is rejected. One method to
deal with this situation would be to repeatedly choose a new position of the two
clusters, starting from their original location and orientation, until
the move step is accepted. This method leads to incorrect simulation
results as is demonstrated for the example of two hard spheres in a
periodic box in Fig.~2. Here a histogram of the distance distribution
between the two spheres, scaled by the volume of a spherical shell, is
shown in a range in which boundary effects are not observed. By
choosing a new position until successful acceptance of the move step
(green histogram), an effective repulsion between the two spheres is
introduced, leading to a depletion zone at small particle distances
instead of the expected uniform distribution. If, instead of choosing
a new position, we set the two clusters involved in a steric collision
back to their original position (and orientation) before the move
step, we recover the expected uniform equilibrium distribution (shown
in red in Fig.~2). Thus, this method ensures that the correct
equilibrium distribution is realized.

\subsection{Local rules}
In our approach, clusters are rigid assemblies consisting of
one or multiple proteins. Assuming that the assembly geometry is determined
by unique local interactions, a set of local rules is necessary to describe the 
new relative orientation and position of two reacting clusters.\cite{Berger1994}
These local rules are encoded in the bond that is formed between the clusters.
Each bond carries the information of the desired center-to-center (ctc) vector
of the two proteins directly involved in the binding process as well
as their relative orientation (torsion vectors). Upon reaction the two
clusters instantaneously flip into the desired relative
configuration. This rule follows from the assumption that the short-ranged forces involved in
the final binding step lead to a fast rearrangement on the time scale
of our BD simulation. The reorientation process is schematically shown
in Fig.~1c. In a first step the clusters are shifted and rotated so
that the predefined ctc-vector matches the real ctc-vector. This
procedure enforces the correct position of the two merging clusters
relative to each other. In a next step the relative orientation of the
clusters is corrected by aligning the projection of the two torsion
vectors on a plane perpendicular to the ctc-vector. The necessary
rotation and translation of the clusters is distributed between them
according to their diffusive weights. This means that for a small and
a large partner, essentially only the small partner is moved, as one
expects for physical reasons; for two similarly sized partners, both
are moved to a similar extent. Since all interactions are local
(e.g. local patches or constraints on orientation/torsion), the
encounter complex corresponds to a region in configuration space
surrounding the desired relative position and orientation. If this
region is sufficiently small, the instantaneous flipping into the
“correct” configuration is comparable to the resolution of the BD
simulation. Reorientation of the clusters during the binding process
can lead to a steric overlap either with another cluster or between
the merging clusters. If this is the case, the reaction is rejected
and the old positions and orientations of the clusters before the move
step are resumed. The steric collision of two merging clusters
reflects that during the assembly process incompatible fragments can
prevent the formation of the desired structure. For example if the
desired structure encoded by the local bonds is a ring consisting of
five proteins, two ring fragments each containing three proteins
cannot bind with each other.

\subsection{Microscopic and macroscopic rates}
\begin{figure}
    \includegraphics[width=8.5cm]{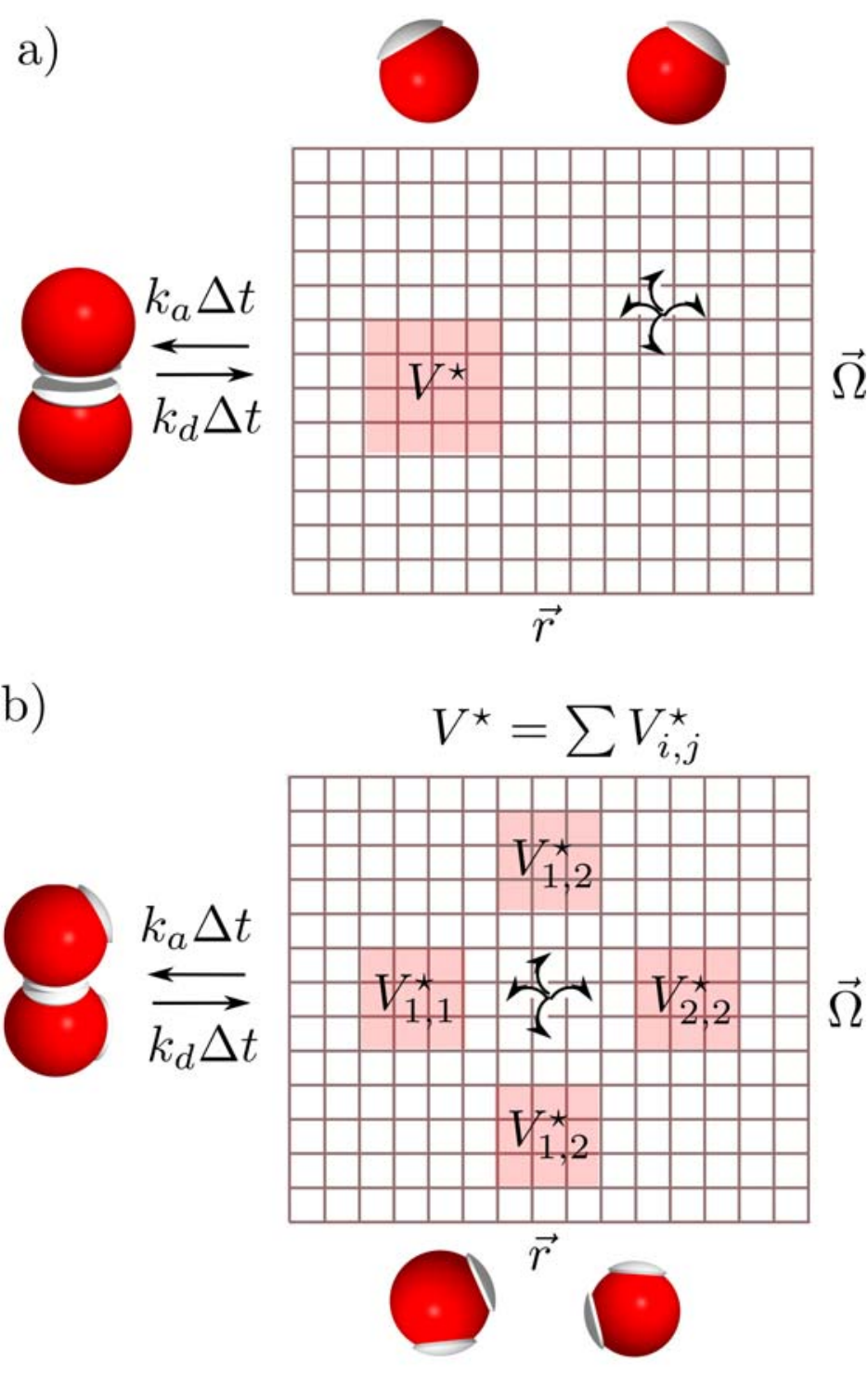}
     \caption{\label{Fig.3} Illustration of the reactive volume $V^\star$. The
configuration space is defined by the relative position $\vec{r}$ of
the particles and the $2\times3$ dimensional orientation vector
$\vec{\Omega}$. a) In the case of one reactive patch $V^\star$
corresponds to the region in configuration space in which two
particles are considered to be in encounter (Eq.~\ref{Eq. Encounter1}
and Eq.~\ref{Eq. Encounter2}). b) For particles with multiple reactive
patches different regions $V^\star_{i,j}$ in configuration space
correspond to an encounter mediated by one particular patch
combination $i$,$j$. For identical microscopic rates $k_a$ and
non-overlapping $V^\star_{i,j}$ the total reactive volume $V^\star$
between the two particles is given by the sum over the encounter
volumes associated with each patch combination.}
\end{figure}

In order to implement full reversibility, both reaction directions are
treated as stochastic reactions with two corresponding microscopic
rates. If the reactive patches of two clusters form an encounter,
specified by the constraints in Eq.~\ref{Eq. Encounter1} and
Eq.~\ref{Eq. Encounter2}, they can react with a bond-specific rate
$k_a^\prime$. Similarly an existing bond can dissociate with a bond-specific
rate $k_d^\prime$. Here we assume that the waiting times for
association or dissociation of a bond between two clusters in
encounter are Poisson-distributed, $P(t,k_i^\prime)=k_i^\prime
\exp(-k_i^\prime t)$, $i=\{a,d\}$. In this case the probability that no
association or dissociation has occurred after a timestep $\Delta t$ is given
by $S(\Delta t,k_i^\prime)=1-\int_{0}^{\Delta
t}P(t,k_i^\prime)dt$ and hence the probabilities for bond
dissociation or bond formation are given by: \cite{Morelli2008}
\begin{align}
 P_\text{assoc}=1-S(\Delta t,k_a^\prime)\approx k_a^\prime \Delta t \text{ ,}\ 
 P_\text{dissoc}&=1-S(\Delta t,k_d^\prime)\approx k_d^\prime \Delta t \text{.}
 \label{Eq. P_bond}
\end{align}
The approximations used in Eq.\ref{Eq. P_bond} are valid if  
$k_d^\prime \Delta t \ll 1$ and $k_a^\prime \Delta t \ll 1$, which is
the case throughout this work.

We now show how the microscopic reaction rates $k_a^\prime$ and $k_d^\prime$ can be
related to macroscopic reaction rates. In a macroscopic
framework the reaction scheme in Eq.~\ref{Eq. encounter}
can be interpreted as a system of ordinary differential equations
describing the changes in the concentrations $c_A$,$c_B$,$c_{A\cdot
  B}$ and $c_C$ of the different species involved in the reaction
. \cite{Eigen1974} In this framework the encounter is understood as a
single, intermediate state connecting the bound complex $C$ and the
unbound $A$ and $B$ particles. The encounter can either react to the
final complex with the first order rate $k_a$ or decay into two
separated particles with the first order rate $k_{D,b}$. An encounter
can be formed by the first order decay of cluster $C$ with a rate
$k_d$ or by the diffusive association of an $A$ and $B$ particle
described by the second order rate $k_D$. Here $k_D$ has the dimension
$\text{m}^3/\text{s}$ if we consider particle concentrations in 3 dimensions.
Using a steady state approximation of the encounter complex, the following
relations are obtained for the overall forward and backward rates and
the equilibrium constant \cite{Eigen1974}:
\begin{align}
  k_+=&\frac{k_D k_a}{k_a+k_{D,b}}  \text{ , }
  k_-=\frac{k_{D,b} k_d}{k_a+k_{D,b}} \label{Eq k+ k-} \\
  \Rightarrow K_\text{eq}=&\frac{k_+}{k_-}=\frac{k_D}{k_{D,b}} \frac{k_a}{k_d}
\text{.}
  \label{Eq Keq2}
\end{align}
In the case of a diffusion-limited reaction $k_a \gg
k_{D,b}$. \cite{Eigen1974} In this case the overall forward and reverse
rates simplify into the purely diffusion-limited forward rate
$k_+\approx k_{D}$ and the backward rate $k_-$ becomes $k_-\approx
k_{D,b} k_d / k_a$. In the case that the reaction is reaction-limited
($k_{D,b} \gg k_a$), the overall forward rate becomes $k_+ \approx
k_D k_a / k_{D,b}$ and the overall backward rate is $k_- \approx
 k_d$.

In the framework of the Fokker-Planck equation two spherical particles
are considered to be in an encounter if they are in
contact. \cite{Smoluchowski1917,Collins1949} Finite reactivity for
particles in encounter was introduced by Collins and
Kimball \cite{Collins1949} in the form of a radiation boundary
condition in which the rate $\kappa_a$ (termed $\kappa$ in
Ref. \cite{Shoup1982} and $k$ in Ref. \cite{Collins1949}) relates the
concentration at contact to the reactive flux. By comparing the escape
probabilities from the encounter in the Fokker-Planck framework and
the macroscopic rate equation framework, Shoup and
Szabo showed that $\kappa_a$ can be related to
macroscopic rates by identifying $\kappa_a=k_D k_a /k_{D,b}$ and
$K_\text{eq}=\kappa_a/k_d$. \cite{Shoup1982} Agmon and Szabo derived
the same relation for the equilibrium constant
($K_\text{eq}=\kappa_a/k_d$) by considering an isolated pair of
reactive spheres using a back-reaction boundary
condition. \cite{Agmon1990} 

Here we show how the microscopic reaction rate used to describe
reactions in our simulation approach can be related to the reaction
scheme in Eq.~\ref{Eq. encounter} and to the equilibrium constant
(Eq.~\ref{Eq Keq2}). Inspired by the work of Shoup and Szabo
\cite{Shoup1982} we calculate the equilibrium constant for the
formation of the encounter. In our approach the encounter is defined
as a region in configuration space (Eq.~\ref{Eq. Encounter1} and
Eq.~\ref{Eq. Encounter2}) around the desired relative position in the
bound complex instead of a boundary as commonly used in the
Fokker-Planck picture. The configuration space of two rigid unbound
particles in a periodic box of volume $V$ can be described by their
relative position vector $\vec{r}$, their center-of-mass vector
$\vec{R}$ and $2\times3$ angular coordinates $\vec{\omega}_1$ and
$\vec{\omega}_2$. Assuming free diffusive motion without reactions all
configurations in the two particle configuration space are equally
probable and they only interact by steric repulsion. By using the thermodynamic extremum
principle for the Gibbs free energy, we can relate the equilibrium
constant for the formation of an encounter to the partition sums of
the free molecules $z_A$ and $z_B$ and to the partition sum of the
encounter complex $z_{A\cdot B}$ \cite{Hill1986,Alsallaq2007a}:
\begin{align}
 K_\text{eq}^\text{enc}=&\frac{k_D}{k_{D,b}}=\frac{(z_{A\cdot B}/V)}{(z_A/V) (z_B/V)} \label{Keq encounter 1} \\
 z_A=&z_B= \int_{V} d^3x \int d^3\omega = V \times 4 \pi \times 2 \pi   \\
 z_{A\cdot B}=& \int_{V} d^3 R \int d^3r \prod_{i=1}^2\big(\int d^3\omega_i) \label{Keq encounter 1.2} \\
 \Rightarrow K_\text{eq}^\text{enc}=&\int d^3r 
\prod_{i=1}^2\big(\frac{1}{8\pi^2} \int d^3\omega_i \big)=:V^{\star} \text{.}
 \label{Keq encounter 2}
\end{align}
For the single molecules all positions and orientations (neglecting the
excluded volume) are accessible, resulting in a factor of $V$ from the
integration over the translational degrees of freedom and a factor of
$8\pi^2$ from the orientational degrees of freedom of the rigid
molecules. To determine the partition sum of the encounter complex
$z_{A\cdot B}$, we first integrate out the center-of-mass coordinate
$\vec{R}$ of the two molecules resulting in a factor of $V$. The
boundaries of the remaining integrals over the relative position
coordinate $\vec{r}$ and the orientation coordinates $\vec{\omega}_1$
and $\vec{\omega}_2$ are determined by the patch definition and the
particle geometries. They follow from Eq.~\ref{Eq. Encounter1} and
Eq.~\ref{Eq. Encounter2} and have to be calculated for every specific
case, as will be discussed in more detail below. The resulting
reactive volume $V^\star$ is the central concept to relate
microscopic and macroscopic rates. From Eq.~\ref{Keq encounter 2}
we see that it equals the equilibrium constant $K_\text{eq}^\text{enc}$.

If we now allow for reactions, not every configuration in the
encounter region is equally populated as some particles will already
react before they explore the inner region of $V^\star$. However, if
on average the encounter region is well explored before an
association, we can approximate the real distribution within $V^\star$
by a uniform distribution.  This approximation is equivalent to the
underlying assumption in the work of Pogson et al \cite{Pogson2006}
and Klann et al. \cite{Klann2011} who introduced a reactive volume
(and an intrinsic association rate \cite{Klann2011}) based on the
assumption that the particles are randomly distributed in the box at
all times. 

Alsallaq and Zhou also used thermodynamic arguments to determine
the equilibrium constant for the bound complex. \cite{Alsallaq2007a}
However, in contrast to this work we consider all configurations
within the generalized reactive volume $V^\star$ as encounter
configurations. By introducing finite reactivity with the microscopic
reaction rate $k_a^\prime$ with which a bound complex can be formed
from the encounter region (see Fig.~3a), our approach can account for
reactions which are reaction- or diffusion-limited, as will be shown
below in the results section. To study assembly, this is of great
importance as intermediates emerging during the assembly process can
have very different diffusion properties and a reaction that was
reaction-limited for small clusters can become diffusion-limited for
larger clusters.
   
Thus, by identifying the equilibrium constant for the formation of the
encounter with the reactive volume $V^\star$
(Eq.~\ref{Keq encounter 2}) and relegating bond association and dissociation to the
reaction rates, we can identify the reaction rates
$k_a^\prime$ and $k_d^\prime$ in the microscopic framework with the
corresponding reaction rates $k_a$ and $k_d$ in the macroscopic
framework (Eq.~\ref{Eq Keq2}) and the equilibrium constant is given
by:
\begin{align}
K_\text{eq}=&\frac{k_D}{k_{D,b}} \frac{k_a}{k_d}=V^\star k_a/k_d\ .
\label{Eq. ka_prime}
\end{align}
In accordance with the work by Shoup and Szabo \cite{Shoup1982} the rate
$\kappa_a=V^\star k_a$ is given by the product of the equilibrium
constant for the formation of the encounter complex and the reaction
rate for the formation of the final complex. This rate can be
identified with the reaction rate used by Morelli and ten
Wolde \cite{Morelli2008} ($k_a$ in ref. \cite{Morelli2008}).

While the reactive volume $V^\star=k_D/k_{D,b}$ is
sufficient to predict the correct equilibrium constant
(Eq.~\ref{Eq. ka_prime}), the kinetics of the reactions depend on the
absolute values of the diffusive rates $k_D$ and $k_{D,b}$. Thus in order
to determine the macroscopic rates $k_+$ and $k_-$ defined in
Eq.~\ref{Eq k+ k-}, we need to evaluate $k_D$. This can be done within
our simulation approach based on an algorithm by Zhou
\cite{Zhou1990}. In this algorithm $k_D$ can be calculated from the
survival probability $S(t)$ of two particles starting in encounter by
\cite{Zhou1990,Alsallaq2007}:
\begin{equation}
 k_D=\lim_{t\rightarrow \infty} \kappa_a \frac{S(t)}{1-S(t)} \text{.} \label{kD
Zhou}
\end{equation}
Given the diffusive on-rate $k_D$, the diffusive off-rate $k_{D,b}$
can be simply calculated by $k_{D,b}=k_D/V^\star$ (Eq.~\ref{Keq encounter 1}).

Any simulation algorithm with underlying reversible dynamics needs to
satisfy detailed balance. To establish detailed balance we follow the
approach by Morelli and ten Wolde \cite{Morelli2008} who inferred from
a detailed balance consideration that the relative position
distribution of two spherical particles prior to a reaction has to be
identical (after renormalization) to the position distribution of the
two particles after dissociation. For two hard spheres covered with a
reactive shell, it has been argued
that detailed balance can be introduced by placing the two hard
spheres according to a radial uniform distribution within the reactive
shell followed by one additional move step \cite{Paijmans}.
Here we generalize this idea as follows: Upon the dissociation of two
clusters, a new configuration is chosen uniformly within the encounter
region $V^\star$ followed by a diffusive move step of the two
clusters.  To establish the uniform distribution in $V^\star$ we
exploit the time invariance symmetry of the diffusion propagator.
This symmetry ensures that in a confined
volume of the configuration space a uniform distribution is
established as the steady state distribution. As $V^\star$ describes a
configuration around the desired relative configuration of the
proteins, we can generate a uniform distribution without prior
knowledge of the exact shape of the reactive volume by performing
various ``pseudo-diffusion'' steps of the two clusters starting from their
predefined relative configuration. These ``pseudo-diffusion'' steps are
confined to the region $V^\star$ in the relative accessible
configuration space of the two partners. We call this procedure
``pseudo-diffusion'' as the two dissociating clusters are propagated
diffusively within $V^\star$, however, the timestep used to establish
the uniform distribution has no physical meaning. This procedure for
generating a uniform distribution within the reactive area does not
depend on the particle or patch geometry. Thus it is especially suited
to study assembly dynamics as parts of the reactive volume might
become sterically blocked during the assembly process. These steric
effects are automatically taken into account with our method. The
effect of steric blocking and its consequences for assembly will be
discussed in the results section.  Moreover, this method would also
remain valid when including long range electrostatic interactions. In
this case the distribution within the encounter complex in equilibrium
would not be uniform. However, the ``pseudo-diffusive'' motion step
would lead to the expected distribution within $V^\star$.

In general the encounter volume $V^\star$ has to be understood as the
volume of all two particle configurations resulting in an encounter
between two clusters (see Fig.~3b). This means that for clusters with multiple
patches the total encounter volume $V^\star=\bigcup V^\star_{p_i,p_j}$
where $V^\star_{p_i,p_j}$ is the encounter volume between a pair of
patches. In the case of a dissociation of clusters with multiple
patches a uniform configuration in the subvolume $V^\star_{p_i,p_j}$
of the patches that formed the bond is realized. If the different
patches have a different microscopic reactivity, the subvolumes have
to be treated separately and the equilibrium constant for the complex
formation is given by the sum of the different equilibrium constants
for different bonds.

\subsection{Intrinsic association}

Proteins with multiple reactive patches can
assemble in structures containing loops. In this case an already
connected cluster can contain open bonds with two patches being in
permanent encounter due to the geometry of the cluster. These
patches can bind with the internal association rate
$k_{a}^\text{intra}$, where we again assume Poisson-distributed waiting times
for the bond formation. The bond formation in this case is
fundamentally different from the above discussed process as no diffusion is
involved. Since clusters are rigid objects, internal bond formation
does not change the shape of the cluster. Thus it can be considered as
an internal process stabilizing an already existing structure. Given a
dissociation rate of $k_d$ for a specific bond, we can relate the
internal association rate to the free energy $E$ of bond
formation. For a closed loop containing $n$ bonds, there are $n$
different configurations with one open bond in the loop and the
fraction of the open and closed loop in equilibrium is given by:
  \begin{align}
  \frac{p_\text{open}}{p_\text{closed}}=&\frac{n
k_d}{k_{a}^\text{intra}}=\frac{Z_\text{open}}{Z_\text{closed}}=\frac{n}{e^{
-\beta E}} \Rightarrow k_{a}^\text{intra}=k_d e^{-\beta E} \text{.} \label{Eq.
internal} 
\end{align}
Thus, the internal association regulates the fraction of open and
closed loop configurations.
In a more realistic scenario, the dissociation of one bond in a loop will
enable an enhanced internal movement. This will be reflected in an
increase in the entropy which would shift the ratio of open and closed
ring structures towards the open state (smaller
$k_{a}^\text{intra}$). In principle it should be possible to calculate
the change in energy and entropy upon dissociation of a bond using detailed
MD simulations. Such simulations could be used to specify the internal
association rate.

\subsection{Outline of the algorithm}

In this part we will describe the work flow of our algorithm. Once a
starting configuration is seeded the following steps are repeated
iteratively. Firstly every particle is translated and rotated
according to its diffusive properties in the framework of free
Brownian motion within a timestep $\Delta t$. With the new positions
and orientations all clusters with steric overlaps as well as all
clusters participating in a reaction are tagged. A reaction between
two clusters occurs with the bond specific probability $k_a \Delta t$
if they are in encounter. In a next step all steric overlaps
(including box overlaps if the box is non-periodic) are corrected by
setting the involved particles back to their original position before
the move step. This might lead to further steric overlaps as a higher
order effect. These overlaps are then also corrected. In this step
clusters tagged for a reaction click into their predefined relative positions
and orientations (see Fig.~1c). If a reaction leads to a steric overlap
either between the two merging clusters or with other clusters, the
reaction is rejected and the particles reassume their configuration
before the move step. Collisions with other clusters after the
reorientations step is a higher order effect in the concentration of
the particles. In dilute systems, especially for reactive
configuration volumes around the desired relative configuration, this
happens rarely.  In a next step open, internal bonds can be closed
with probability $P^\text{acc}_\text{intra}=k_{a}^\text{intra} \Delta
t$. Finally each existing bond can dissociate with the bond specific
probability $k_d \Delta t$. If this leads to two unconnected cluster
fragments, the diffusive properties of the fragments are
calculated. Subsequently a uniform distribution within the encounter
volume $V^\star$ is realized by ``pseudo-diffusive'' motion preserving the
encounter followed by one unconstrained diffusion step of the two
clusters. All simulations have been performed at room temperature
$T=293 K$ and with the viscosity of water
$\eta=1\text{mPas}$.

\section{Results and Discussion}

\subsection{The reactive volume}
\begin{figure}
     \includegraphics[width=8.5cm]{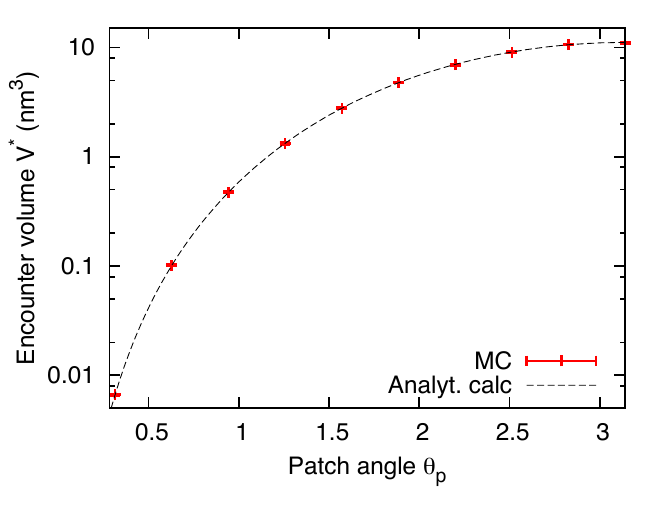}
     \caption{\label{Fig.4} $V^\star$ for two identical hard spheres of radius $R=1\text{nm}$
equipped with one polar patch of radius $R_p=1.1$nm as a function of
the patch angle $\theta_p$. The analytical result from
Eq.~\ref{Eq. V Kern-Frenkel} is shown with a dashed line while the MC
estimates (according to Eq.~\ref{Eq. MC}) are represented by red
points with error bars.}
\end{figure}

As discussed in the previous section, we can relate the microscopic
rates to the macroscopic equilibrium constant using the concept of a
generalized reactive volume $V^\star$ in configuration space
(Eq.~\ref{Eq. ka_prime}). Here we introduce a model system in which it
is possible to analytically calculate $V^\star$ and show that the
results from our MC scheme (Eq.~\ref{Eq. MC}) agree well with the
analytical results.  In this model system a protein is described by
one hard sphere of radius R equipped with one polar patch as depicted
in Fig.~3a. The patch is centered at the origin of the protein. It has
a radius of $R_p$ and an opening angle of $\theta_p \in [0,\pi]$
around the z-axis of the proteins. With our definition of the
encounter complex for two such particles (Eq.~\ref{Eq. Encounter1} and
Eq.~\ref{Eq. Encounter2}), we can analytically calculate $V^\star$ by
decomposing the radial and orientational part, similar to the potential model
proposed by Kern and Frenkel\cite{Kern2003}:
\begin{align}
 V^\star=&V^\star_\text{rad} \times V^\star_\text{ori}  \label{Eq. V Kern-Frenkel}\\
 V^\star_\text{rad}=&\frac{4}{3}\pi((R^p)^3-R^3) \text{, }
 V^\star_\text{ori}=\big(\frac{1}{4\pi}\int_0^{2\pi}d\phi \int_0^{\theta_p}
\sin(\theta)d\theta\big)^2=\frac{1}{4}(1-\cos(\theta_p))^2 \text{.}  \nonumber
\end{align}
The square in the orientation part arises from the fact that we have
to consider two independently rotating particles and only if the
orientation of both particles is correct, an encounter is reached (see
Eq.~\ref{Eq. Encounter2}). For $\theta_p=\pi$ the orientation
constraint vanishes and the spherical result is recovered.

As analytically calculating the reactive volume is only feasible for
some special geometries, it can be numerically pre-calculated for the
elementary assembly blocks using a MC integration
scheme. In this scheme we sample different configurations by randomly
positioning two clusters with random orientations in a periodic box of
volume $V_\text{box}$. We repeat this procedure $N$ times and can
estimate $V^\star$ by counting the fraction of trials
$n=N_\text{encounter}/N_\text{total}$ that result in an encounter
configuration. Then we simply have
\begin{equation}
V^\star_{MC} = n V_\text{box} \text{.} \label{Eq. MC}
\end{equation}
In Fig.~4 the configuration space volume $V^\star$ for two identical,
spherical proteins equipped with a polar patch ($R=1$nm and
$R_p=1.1$nm) is shown as a function of the opening angle
$\theta_p$. Comparing the analytical result (black line) and MC
simulation (Eq.~\ref{Eq. MC}) we see that our simple MC scheme
correctly estimates $V^\star$. For proteins equipped with multiple,
non-overlapping patches which can bind with each other via different
patch combinations, we have to distinguish between the encounter
volume $V^\star_{p_i,p_j}$ of two specific patches $p_i$ and $p_j$ and
the total encounter volume $V^\star$ of two proteins. The total
configuration volume $V^\star$ is given by the union of all volumes
$V^\star_{p_i,p_j}$ for the encounter of two different patches $i$,$j$
that can bind to each other. If the encounter volumes of different
patch combinations are disjoint, the total configuration space volume
for the encounter of two clusters is given by summing over all
$V^\star_{p_i,p_j}$. For multiple, identical patches that can bind
with each other, the total reactive volume $V^\star$ is given by
multiplying the configuration volume $V^\star_{p_i,p_j}$ of one patch
combination with a combinatorial prefactor accounting for the
multiplicity of the different patch configurations if the
corresponding volumes are not overlapping in configuration space.

\subsection{Bimolecular reactions}
\begin{figure}[!H]
       \includegraphics[width=8.5cm]{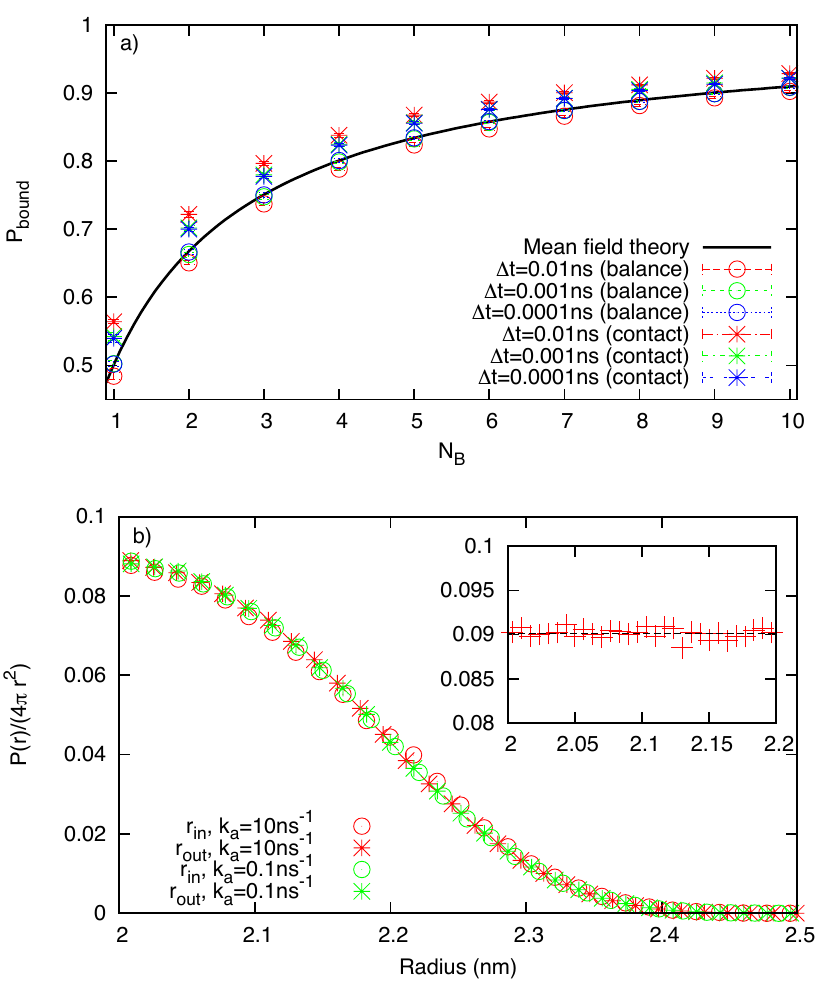}
      \caption{\label{Fig.5} Equilibrium bound probabilities and radial distribution for
spherically reactive particles.  a) The probability of one $A$ particle
being bound to a $B$ particle is shown as a function of the number of $B$
particles according to Eq.~\ref{Eq. MF} (black line) for
$K_\text{eq}=V_\text{box}$. The following parameters have been used
for the simulations: $R_A=R_B=1$nm, $R_{p,A}=R_{p,B}=1.1$nm,
$\theta_{p,A}=\theta_{p,B}=\pi$ leading to
$V^\star\approx=11.09\text{nm}^3$ (Eq.~\ref{Eq. V Kern-Frenkel}). For a
given $k_a=10\text{ns}^{-1}$, the dissociation rate $k_d$ is chosen so
that $V^\star\frac{k_a }{k_d}=V_\text{box}=8000$nm$^3$
(Eq.~\ref{Eq. ka_prime}). Circles depict the simulation results
obtained with our algorithm for different time resolutions while stars
correspond to simulations in which detailed balance is violated by
placing particles in contact after dissociation. b) The normalized
radial distribution $\Delta t$ before reaction (circles) and after
dissociation including the additional unconstrained move step (stars)
is shown for two different association rates $k_a=10$ns$^{-1}$ (red)
and $k_a=0.1$ns$^{-1}$(green) with a time resolution of $\Delta
t=0.01\text{ns}$. The inset shows the comparison of the uniform radial
distribution established by ``pseudo-diffusive'' motion constraint to
$V^\star$ (red points) and the expected uniform distribution in a
spherical shell (black line).  }
\end{figure}
To verify our algorithm and to show the importance of detailed balance,
we first analyze a system with a small number of proteins only. In
detail we consider a situation in which one particle of type $A$
and $N_B$ particles of type $B$ are placed in a periodic box of volume
$V_\text{box}$. In this case we can relate the concentration $c_C$ of
the complex $C$ to the fraction of time $t_b$ in which particle $A$ is
bound to particle $B$ ($p_b=t_b/(t_u+t_b)$), while the concentration
$c_A$ of particle $A$ can be related to the fraction of time $t_u$ in
which $A$ is unbound ($p_b=t_u/(t_u+t_b)$). The unbound $A$ particle
is surrounded by a concentration of $c_B=N_B/V$ particles of type
$B$. Here $V=V_\text{box}-V_\text{excl}$ is the freely accessible
volume. For proteins of identical radius ($R_A=R_B=R$) the excluded
volume can be approximated by $V_\text{exc}=4\pi N_B(2R)^3/3$. Thus we
can relate the probability that $A$ is bound to the equilibrium
constant of the reaction by: \cite{Agmon1990,Morelli2008}
\begin{align}
 &K_\text{eq}=\frac{c_C}{c_B
c_A}=\frac{p_\text{bound}}{\frac{N_B}{V}\big(1-p_\text{bound}\big)} 
  \Rightarrow p_\text{bound}(N_B)=\frac{K_\text{eq}N_B}{K_\text{eq}N_B
+V}\text{.}
\label{Eq. MF}
\end{align}
Eq.~\ref{Eq. MF} follows from elementary thermodynamic arguments and is
valid irrespective of the details of the binding interactions. In the
following subsections we will show that the correct equilibrium bound
probability is reproduced with our algorithm for spherically reactive
and patchy particles when appropriately taking the generalized
encounter volume $V^\star$ into account. We also show that detailed
balance is essential to reproduce $p_\text{bound}$ from Eq.~\ref{Eq. MF}.
 
\subsubsection{Spherically reactive particles}

Here we compare our BD algorithm with the mean field result
(Eq.~\ref{Eq. MF}) for the case of spherically reactive $A$ and $B$
particles ($R_A=R_B=1$nm, $R_{p,A}=R_{p,B}=1.1$nm,
$\theta_{p,A}=\theta_{p,B}=\pi$, $V^\star=11.09$nm$^3$) and show that
our algorithm fulfills detailed balance in this case. Given an
association rate of $k_a=10$ns$^{-1}$ we chose $k_d$ so that
$K_\text{eq}=V^\star\frac{k_a }{k_d}=V$. In Fig.~5a the probability of
particle $A$ being bound to particle $B$ is shown as a function of the
number of $B$ particles $N_B$ and compared to the mean field result
(black line). Placing particles uniformly in the reactive shell by
``pseudo-diffusive'' motion and allowing for one unconstrained move step
of the two proteins involved in the dissociation according to the
algorithm described above (circles) leads to very good agreement
between the mean field theory and the simulation results. Only for a
very large timestep $\Delta t=0.01\text{ns}$ leading to a large
reaction probability of $0.1$ a small deviation between the simulation
results and the expected mean field theory is observed. On the
contrary, violating detailed balance by placing particles in contact
after the dissociation (stars) leads to an overestimation of
$p_\text{bound}$ and the simulation results cannot be reconciled with
the mean field theory. Although reducing the timestep from $\Delta t
=0.01\text{ns}$ to $\Delta t =0.001\text{ns}$ leads to a modest
improvement for the simulation results with contact dissociation, no
further improvement is observed when decreasing the timestep
further. This shows that detailed balance is essential in the dissociation
step for the agreement between simulation results and macroscopic theory. These
findings agree with those obtained by Morelli and ten Wolde who performed a
similar simulation to test their algorithm. \cite{Morelli2008} 

In Fig.~5b we verify that our algorithm indeed obeys detailed balance
by comparing the radial distribution of particles before a
reaction (circles) and after dissociation (stars) for two different
association rates. Irrespective of the choice of $k_a$ we observe the
same radial distribution before reaction and after dissociation and
thus our algorithm satisfies detailed balance for this quasi
one-dimensional case. In the inset of Fig.~5b we show a histogram of
the distance between two dissociating particles after the
``pseudo-diffusive'' motion constraint to $V^\star$ without the
additional, unconstrained move step of the two particles. Here we see
that we are indeed able to generate a uniform distribution within
$V^\star$ by exploiting the symmetry of the diffusion propagator as
the histogram of the simulated positions agrees very well with the
expected uniform distribution.
 
\subsubsection{Patchy Particles}
\begin{figure}
     \includegraphics[width=8.5cm]{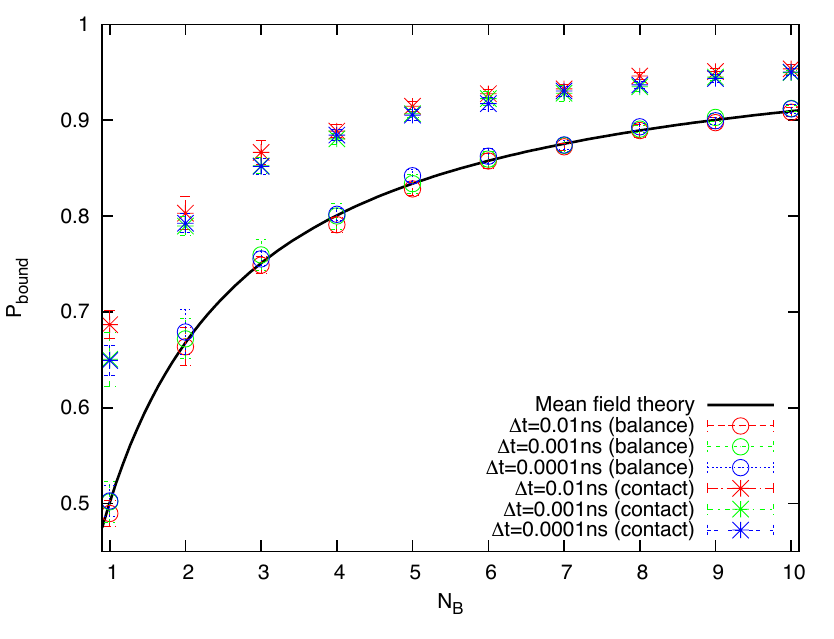}
    \caption{\label{Fig.6} Equilibrium bound probabilities for patchy particles. The
probability of one $A$ particle being bound to a $B$ particle is shown as a
function of the number of $B$ particles according to Eq.~\ref{Eq. MF}
(black line) for $K_\text{eq}=V_\text{box}$. The mean field theory is
independent of the specific particle details. The following parameters
have been used for the simulations: $R_A=R_B=1$nm,$R_{p,A}=R_{p,B}=1.1$nm,
$\theta_{p,A}=\theta_{p,B}=\pi/4$ leading to
$V^\star\approx=0.23\text{nm}^3$ (Eq.~\ref{Eq. V Kern-Frenkel}). For a
given $k_a=10\text{ns}^{-1}$ the dissociation rate $k_d$ is chosen so
that $V^\star \frac{k_a}{k_d}=V_\text{box}=8000$nm$^3$. Circles show
the results of our algorithm for various timesteps while stars
correspond to simulations in which detailed balance is violated by
placing particles in contact and with perfect alignment of the patches
after dissociation.}
\end{figure}

\begin{figure}
     \includegraphics[width=8.5cm]{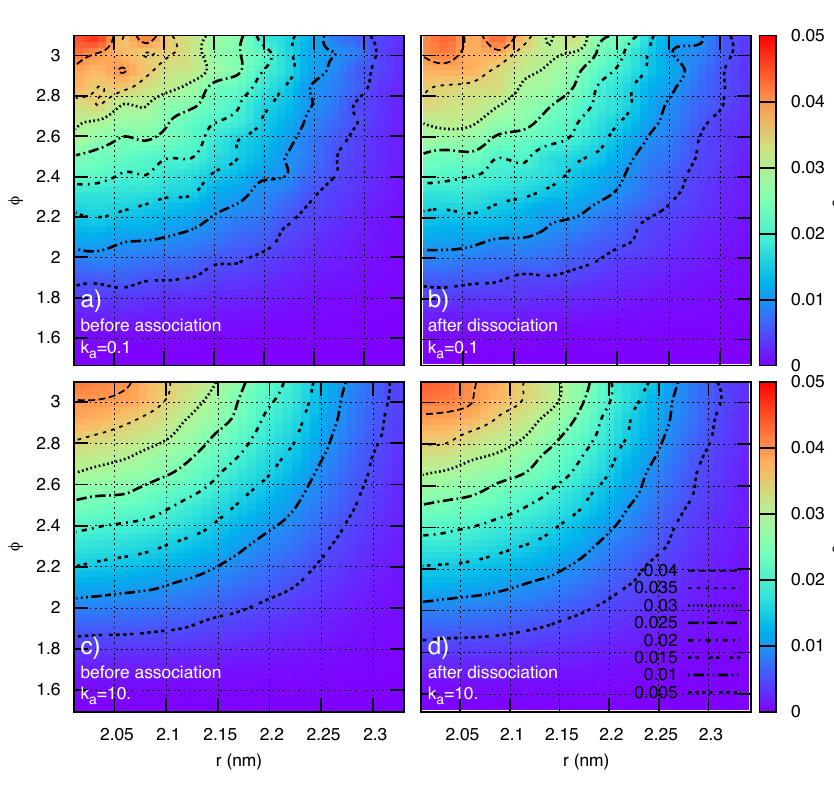}
    \caption{\label{Fig.7} Color-coded histogram of the particle positions before
reaction and after dissociation as a function of the relative distance
$r$ and the angle between the orientations of the two patches $\phi$. 
This histogram has been weighted by the generalized bin volume ($4\pi r^2 dr \times
2 \pi\sin(\phi) d\phi $) . Several contour lines are shown for better
comparability.  In a) and c) the distribution of particle positions
before a reaction is shown for $k_a=0.1$ns$^{-1}$ and
$k_a=10$ns$^{-1}$, respectively. In b) and d) the corresponding
distributions of particle positions after dissociation is shown. The
distributions have been recorded with a timestep resolution of $\Delta
t=0.01\text{ns}$.}  
\end{figure}

After verifying that our algorithm reproduces the expected results for
spherically reactive particles, we will now investigate the effect of
anisotropic reactivity and show that our generalized definition of the
reactive volume enables us to compare our simulation results with the
mean field theory, which is the same as for spherically reactive particles.
Here we consider the same setup as in the previous
part with one particle of type $A$ and $N_B$ particles of type $B$ in
a periodic box ($R_A=R_B=1$nm and $R_{p,A}=R_{p,B}=1.1$nm). This time,
however, both particles are equipped with a polar patch
($\theta_p=\pi/4$) allowing only for an association of two particles
if the patches are sufficiently aligned with the ctc-vector of the
proteins (Eq.~\ref{Eq. Encounter2}). In this case the
reactive volume reduces to $V^\star=0.23$nm$^3$ according to
Eq.~\ref{Eq. V Kern-Frenkel}. Given the reactive volume $V^\star$ and
the association rate $k_a=10$ns$^{-1}$, we again choose $k_d$ in such
a way that $K_\text{eq}=V$. The comparison between the simulation and
mean field theory is shown in Fig.~6. By using the appropriate scaling
with the generalized volume $V^\star$ for the patchy particles, we
observe perfect agreement of the mean field results from
Eq.~\ref{Eq. MF} (black line) with the simulation results when placing
the particles according to the proposed detailed balance algorithm
(circles). When placing the particles in contact (alignment of patch
orientation with ctc-vector and radial contact), we drastically
overestimate the bound probability (stars). Compared to the spherical
case (Fig.~5) this overestimation of $p_\text{bound}$ is much
stronger. Thus, in the case of localized reactivity it is even more
important to carefully consider the relative position and orientation
of dissociating proteins. Moreover our results confirm the relation
between microscopic rates and the macroscopic equilibrium constant
given in Eq.~\ref{Eq. ka_prime}.

In Fig.~7 we again show the distribution of the relative position and
orientation of two particles equipped with a polar patch before a
reaction and after dissociation. Here $\phi$ corresponds to the angle
between the orientation vectors of the patches and $r$ is the distance
between the centers of the two particles. Due to the symmetry of the
particles around the z-axis these two parameters provide a good
description of the relevant configuration space. The normalized
probability distribution $p(r,\phi)/(4\pi r^2 \times 2\pi sin(\phi))$
is shown color-coded for two microscopic association rates
$k_a=0.1$ns$^{-1}$ (Fig.~7 upper part) and $k_a=10$ns$^{-1}$ (Fig.~7
lower part) at a time resolution of $\Delta t=0.01$ns$^{-1}$. In the
left part of Fig.~7 the distribution before reaction is shown while in
the right part the corresponding distribution after dissociation is
depicted.  As expected the distribution has its maximum for particles
in contact and anti-parallel aligned patch orientations ($r=2.0$nm and
$\phi=\pi$). Although the distributions fluctuate more for a smaller
association constant (Fig.~7a and 7b), as can be seen by looking at the
contour lines, very good agreement between the distribution before
association and after dissociation is observed for both association
parameters. This shows that our algorithm maintains detailed balance
in this case of non-spherical reactivity. Changing the association
rate by two orders in magnitude does not affect the distributions
suggesting that our algorithm works well for a large spectrum of
microscopic parameters. This first example of non-spherically reactive
particles shows that it is essential to appropriately scale the
macroscopic parameters using the concept of a generalized reactive
volume in configuration space. Moreover, careful consideration of the
relative position and orientation of the dissociating particle is
necessary to obtain quantitative agreement between BD simulations and
mean field theory or experimental results. 

\subsection{Beyond bimolecular reaction: Assembly of a pentameric ring}
\begin{figure}
   \includegraphics[width=8.5cm]{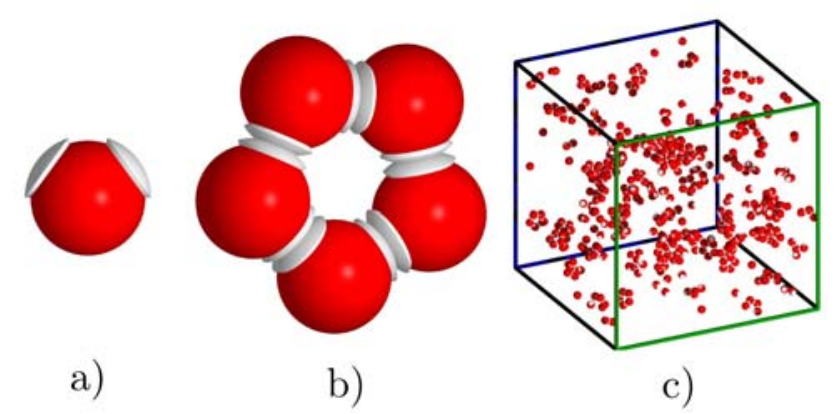}
  \caption{\label{Fig.8} Simulation of the assembly of a pentameric ring structure.
a) Elementary spherical building block of radius $R=1$nm for the
pentameric ring equipped with two patches. The angle between the
centers of the patches is determined by the desired ring
structure. Each patch has an opening angle of $\theta_p=\pi/5$ and a
radius of $R_p=1.1$nm. b) Fully assembled ring structure. c)
Simulation snapshot of a typical box containing 500 proteins.}  
\end{figure}

In this section we analyze the assembly of a pentameric ring structure
consisting of 5 proteins (Fig.~8b) as an example of an assembly process
involving more than two partners. As elementary building blocks we
consider spherical proteins of radius $R=1$nm, each equipped with two
reactive patches (Fig.~8a). The patches are centered at the origin of
the protein and both have a radius of of $R_p=1.1$nm and an opening
angle of $\theta_p=\pi/5$. The geometry of the ring is reflected in
the different orientation vectors $\vec{o}$ associated with each
patch, and the relative position and orientation of the proteins are
encoded in the bond structure.

Here we develop a rate equation approach based on a set of ordinary
differential equations (ODEs) to describe the changes in concentration
of the different fragments sizes. We demonstrate how the macroscopic
reaction rates can be calculated from the microscopic reaction
parameters and diffusive properties of the intermediates. The
parameter free comparison of our simulation results with the rate
equation predictions shows excellent agreement. This demonstrates that
we are indeed able to predict macroscopic reaction rates from our
simulations taking into account changes in diffusive properties of the
assembly intermediates as well as steric effects arising during the
assembly process.

\subsubsection{Macroscopic rates}
\begin{table}[h!]
	      \begin{tabular}{|c|c|c|c|c|}
		    \hline
		    Size  
		    &
		    1
		    &
		    2
		     &
		    3
		     &
		    4 \\
		    \hline
		    1
		    & 0.40 & $0.40$	& $0.40$	& $0.24$\\
		    \hline
		    2
		    &$0.40$	&$0.40$	&$0.36$	& - \\
		    \hline
		    3
		    &$0.40$	&$0.36$	&-	&- \\
		    \hline
		    4
		    &$0.24$	&-	&-	&- \\
		    \hline
	      \end{tabular}
	      \caption{\label{Tab.1} Generalized reactive volumes
                $V^\star_{i,j}$ in $\text{nm}^3$ for the encounter of
                different ring fragments calculated by MC sampling
                according to Eq.~\ref{Eq. MC} and analytically
                according to Eq.~\ref{Eq. V Kern-Frenkel} for
                $V^\star_{1,1}$, respectively.}
	\end{table}
	
	  \begin{table}[h!]
	      \begin{tabular}{|c|c|c|c|c|}
		    \hline
		    Size
		    &
		    1
		    &
		    2
		     &
		    3
		     &
		    4\\
		    \hline
		    1
		    &3.29 &2.69	& 2.15	& 1.17\\
		    \hline
		    2
		    &2.69	&2.03	&1.52	& - \\
		    \hline
		    3
		    &2.15	&1.52	&-	&- \\
		    \hline
		    4
		    &1.17	&-	&-	&- \\
		    \hline
	      \end{tabular}
	      \caption{\label{Tab.2}Diffusive on-rates $k_D$ for
                different ring fragments in $\text{nm}^3/\text{ns}$. The rates
                have been calculated from the survival probabilities
                of fragments starting in a encounter according to the
                algorithm proposed by Zhou\cite{Zhou1990} with a
                minimum timestep resolution of $\Delta t
                =0.01\text{ns}$.}
	\end{table}

In Eq.~\ref{Eq k+ k-} the reaction rates $k_+$ and $k_-$ are given in
the case of a bimolecular reaction. We now specify the macroscopic
rates $k_+^{i,j}$ and $k_-^{i,j}$ for the reaction between two ring
fragments with $i$ and $j$ proteins. These rates are given by:
\begin{align}
 k_+^{i,j}&=\overbrace{\frac{k_D^{i,j} k_a}{k_{D,b}^{i,j}+k_a}}^{\tilde{k}_+^{i,j}}  (1-\frac{1}{2}\delta_{i,j}) \label{Eq. rates ring I} \\
 k_-^{i,j}&=\underbrace{\frac{k_{D,b}^{i,j}
kd}{k_{D,b}^{i,j}+k_a}}_{\tilde{k}_-^{i,j}}  (2-\delta_{i,j}) \text{.}
\label{Eq. rates ring II} 
\end{align}
$k_+^{i,j}$ describes the association of two ring fragments of size
$i$ and $j$. As the diffusive properties and the accessible reaction
volume change with the size of the fragments $k_+^{i,j}$ is different
for different combination of fragments, albeit the microscopic
reaction rate $k_a$ remains constant. For the reaction of identical
particles the macroscopic rate is reduced by a prefactor of $1/2$
compared to the simulation rate $k_a$. This reflects the fact that for
reactions of identical components ($A+A\rightarrow C$) the observed
macroscopic rate is by a factor of $1/2$ smaller than the rate used in
the simulation due to the number of collisions being proportional to
$N_A(N_A-1)/2!$. \cite{Andrews2004,Gillespie1977} $k_-^{i,j}$
describes the macroscopic dissociation rate of a cluster of size $i+j$
into two ring fragments of size $i$ and $j$, respectively. To determine
the macroscopic reaction rates given in Eq.~\ref{Eq. rates ring I} and
Eq.~\ref{Eq. rates ring II} we need to calculate the diffusive on- and
off-rates $k_D^{i,j}$ and $k_{D,b}^{i,j}=k_D^{i,j}/V^\star_{i,j}$
which change for different combination of fragments. Here
$V^\star_{i,j}$ is the configuration volume associated with the
reaction of two fragments of size $i$ and $j$.

The reactive volume for two different fragment configurations $i$ and
$j$ can be calculated by MC sampling as has been described
in Eq.~\ref{Eq. MC}. The resulting volumes $V^\star_{i,j}$ for two
fragments of size $i$ and $j$ are shown in Table \ref{Tab.1}. For the
reaction of two monomers the reactive volume $V^\star_{1,1}$ can also
be evaluated by using Eq.~\ref{Eq. V Kern-Frenkel} for the encounter of
one specific pair of patches. Taking into account the multiplicity of
bond interactions $2^2$ results in a total encounter volume of
$V^\star_{11}=0.4$nm$^3$. While fragments not combining into a full
ring ($i+j<5$) have the same configuration volume as the monomeric
fragments, the configuration volume for fragments that can combine
into a full ring is reduced. This reflects the fact that, independent
of the particle geometry, the encounter volume and thus the number of
configurations in this volume remains unchanged, unless parts of the reactive
volume become sterically blocked, which is the case for fragments that can
assemble into the full ring structure. Moreover, for these fragments some
configurations with a simultaneous overlap of different patch combinations
exist. As the encounter volumes for different patch combinations are not
disjoint in this case, the total encounter volume is smaller than the sum of the
subvolumes. These effects are naturally included in our simulation
algorithm and for the simulations only knowledge of the reduced volume
for the basic building blocks $V^\star_{1,1}$ is required. However, in
order to compare our simulation results to a macroscopic rate equation
approach, we need to determine the changes in the encounter volumes
for different assembly intermediates.

Moreover, in order to capture the reaction kinetics with the above
defined rates, we need to determine the diffusive on-rates
$k_D^{i,j}$. In order to calculate the diffusive on-rate $k_D^{i,j}$
we follow a scheme which was originally proposed by
Zhou. \cite{Zhou1990}. Here, two ring fragments of size $i$ and $j$
are initially placed in a random configuration in encounter (e.g. a
random configuration within $V^\star_{i,j}$ is chosen).
Starting from this configuration the two
fragments are propagated diffusively within our simulation
framework. If the two fragments form an encounter, they can react with
the probability $k_a \Delta t$. In this case the run is terminated. By
recording the survival probability $S^{i,j}(t)$ (the fraction of runs
which survive until time $t$) $k_D^{i,j}$ can be estimated according
to Eq.~\ref{kD Zhou} with $\kappa^{i,j}_a=k_a V^\star_{i,j}$. In
order to calculate $S^{i,j}(t\rightarrow \infty)$ we use a microscopic
reaction rate of $k_a=1\text{ns}^{-1}$ and an adaptive
timestep scheme with a minimum timestep of $\Delta
t_\text{min}=0.01\text{ns}$ which ensures that all steric collision and encounters are detected
on the resolution of the minimum timestep. \cite{Zhou1990} The diffusive off-rate for the decay
of a fragment into two smaller fragments of size $i$ and $j$ is given
by $k_{D,b}^{i,j}=k_D^{i,j}/V^\star_{i,j}$. The absolute values for
the different ring fragments are given in Tab. \ref{Tab.2}. As
fragments with $(i+j)>5$ cannot combine with each other due to steric
collisions, the diffusive on-rate in this case is $k_D^{i,j}=0$.

\subsubsection{Macroscopic rate equation approach}

After having defined the macroscopic reaction rates for the
association and dissociation of different ring fragments
(Eq.~\ref{Eq. rates ring I} and Eq.~\ref{Eq. rates ring II}) we will now
introduce a system of ODEs to describe the changes in concentration $c_i$ of a
ring fragment of size $i$.  This description assumes that the system
is homogeneous and well mixed at all times. Stochastic fluctuations
arising from the finite number of proteins are neglected. While for
ring fragments containing less than 5 proteins only one state exists,
two states exist for the full ring: an open ring with 4 bonds and a
closed ring with 5 bonds. The closed ring can be considered as an
internal state as the last bond formation in our framework does not
involve any diffusion and it is connected to the open state by the
internal association rate $k_a^\text{intra}$ (see
Eq.~\ref{Eq. internal}). The concentration of the open state is denoted
by $c_5$ while the concentration of the closed state is denoted by
$c_5^\star$.  The changes in concentration for a ring consisting of
$N$ proteins can be described by the following set of ODEs:
\begin{widetext}
\begin{align}
 \dot{c}_i=&-\sum_{l=1}^{N-i}  k_+^{i,l} (1+\delta_{i,l})c_i c_l
	    -\sum_{\substack{l = 1  \\ l \leq i-l}}^{i-1}  k_-^{l,i-l} c_i
	    -k_a^\text{intra}\delta_{i,N}c_N  \nonumber \\
	    &+ \sum_{\substack{l = 1  \\ l \leq i-l}}^{i-1}  k_+^{l,i-l} c_l c_{i-l}
	    +\sum_{l=1}^{N-i}k_-^{i,l}(1+\delta_{i,l})c_{l+i}  
	    +N\delta_{i,N} k_d c_N^\star \label{Eq. rate_eq0} \\
	   =&-\sum_{l=1}^{N-i}  \tilde{k}_+^{i,l}c_i c_l
	     -\sum_{l = 1}^{i-1}  \tilde{k}_-^{l,i-l} c_i
	     -k_a^\text{intra}\delta_{i,N} c_N \nonumber \\
	     &+\frac{1}{2}\sum_{l = 1}^{i-1}   \tilde{k}_+^{l,i-l} c_l c_{i-l}
	     +2\sum_{l=1}^{N-i}\tilde{k}_-^{i,l}c_{l+i} 
	     +N\delta_{i,N} k_d c_N^\star \label{Eq. rate_eq} \\
 \dot{c}_N^\star=&-Nk_d c_N^\star+ k_a^\text{intra} c_N \text{.} \label{Eq.
rate_eq2} 	     
\end{align}
\end{widetext}
In our case $N=5$. Eq.~\ref{Eq. rate_eq0} shows the different processes
which lead to a change in the concentration of a fragment containing
$i$ proteins. The first three terms lead to a decrease in
concentration. The first term describes the change in concentration
due to the reaction of a fragment of size $i$ with another fragment of
size $l$. If $i=l$ a twofold decrease in concentration $c_i$ is
observed. The second term describes the change in concentration due to
the decay of a fragment of size $i$ into two fragments of size $l$ and
$i-l$.  To prevent the double counting of dissociation events the
constraint $l\leq i-l$ for the summation is used here. In the third
term the internal bond formation from the open to the closed state is
described. Similarly in the last three terms all contributions which
lead to an increase in concentration $c_i$ due to association and
dissociation processes are described.  In Eq.~\ref{Eq. rate_eq2} the dynamics of the
closed pentameric ring state is separately described. In the
derivation of Eq.~\ref{Eq. rate_eq} from Eq.~\ref{Eq. rate_eq0} the
symmetry of the matrices $k_\pm^{i,l}=k_\pm^{l,i}$ has been
exploited. The above described set of ODEs obeys mass conservation
($\sum_i (c_i+\delta_{i,5}c_5^\star) i=\text{const}$). Thus, one of
the concentrations could be eliminated as it is dependent on the other
concentrations.

\subsubsection{Comparison between rate equations and BD simulations}
\begin{figure}
 \includegraphics[width=8.5cm]{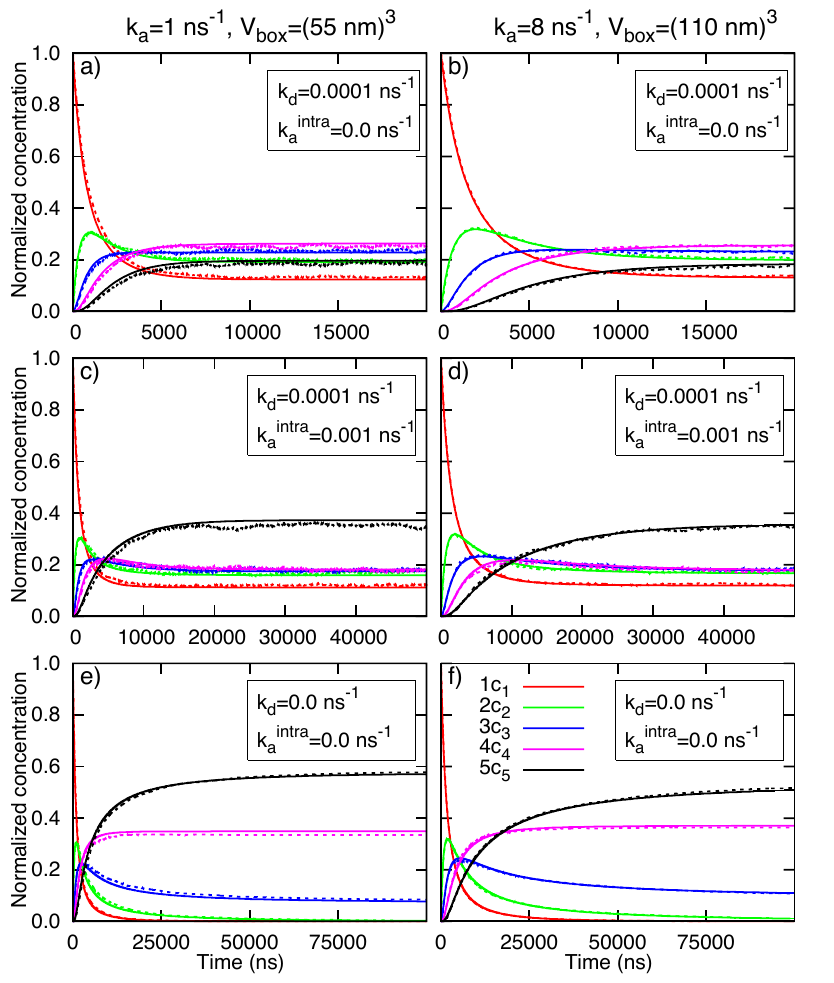}
  \caption{\label{Fig.9} Comparison of the rate equation approach
Eq.~\ref{Eq. rate_eq} (solid lines) and the BD algorithm (dashed
lines). The evolution of the  normalized concentrations of the ring fragments 
$\tilde{c}_i=\frac{ic_i}{c_0}$ with
$c_0=N/V$ are shown for different parameters. All simulations were
started with $N=500$ individual proteins with a time resolution of
$\Delta t=0.01$ns and the simulation results have been averaged over
40 independent trajectories.  The left column corresponds to
reaction-limited assembly with $k_a=1\text{ns}^{-1}$ and
$V_\text{box}=(55\text{nm})^3$. The right column corresponds to
diffusion-influenced assembly with $k_a=8\text{ns}^{-1}$ and
$V_\text{box}=(110\text{nm})^3$ in which the increase in $k_a$ is
balanced by the decrease in concentration. In a) and b) the result for
a reversible reaction with a dissociation rate of
$k_d=0.0001\text{ns}^{-1}$ and without internal bond formation is
($k_{a}^\text{intra}=0.$ns$^{-1}$) shown. In c) and d) the final ring
is additionally stabilized by the formation of an internal bond with
rate $k_{a}^\text{intra}=0.001$ns$^{-1}$ (see
Eq.~\ref{Eq. internal}). In e) and f) the evolution of the cluster size
distribution is shown for irreversible binding
($k_d=0\text{ns}^{-1}$).}
\end{figure} 

After having defined the macroscopic rate equation approach for the
changes in fragment concentration in Eq.~\ref{Eq. rate_eq}, we will now
compare our BD simulation results to the macroscopic rate equation
approach. As all macroscopic rates are fully specified by the
microscopic parameters (Eq.~\ref{Eq. rates ring I} and
Eq.~\ref{Eq. rates ring II}), the comparison between the simulation and
macroscopic theory does not involve any free parameter.  For the BD
simulations we randomly position $N=500$ single proteins in a periodic
box of size $V_\text{box}$ and record the time-course of the fragment
concentrations. All BD simulations were performed at a timestep
resolution of $\Delta t=0.01\text{ns}$ and the resulting
concentrations were averaged over 40 independent trajectories.  For
the comparison with the rate equation approach we use an initial
concentration of $c_0=c_1(t=0)=N/V$. Here $V$ is the accessible
simulation volume which is roughly estimated from the box volume by
$V=V_\text{box}-(N-1)4\pi(2R)^3/3$.

In Fig.~9 the evolution of the normalized fragment concentrations
$\tilde{c}_i=(c_i+\delta_{i,5}c_5^\star) i/c_0$ is
shown. The normalized concentration of a fragment is defined as the
concentration $c_i$ weighted by the number of proteins contained in the fragment
and normalized by the initial concentration.
The normalized concentration predicted by the rate
equation approach is shown with solid lines, while the averaged
simulation results are represented by dashed lines.  In the left
column of Fig.~9 (Fig.~9 a,c and e) a microscopic reaction rate of
$k_a=1\text{ns}^{-1}$ and a box volume of
$V_\text{box}=(55\text{nm})^3$ are chosen.  As the diffusive off-rates
range from $k_{D,b}^{1,1}\approx8.2 \text{ns}^{-1}$ to
$k_{D,b}^{2,3}\approx4.2 \text{ns}^{-1}$, this assembly process can be
considered as reaction-limited ($k_a < k_{D,b}$).  In the case
of a reaction-limited process, the macroscopic off- and on-rates given
in Eq.~\ref{Eq. rates ring I} and Eq.~\ref{Eq. rates ring II} simplify
to $\tilde{k}_+^{i,j}\approx k_D^{i,j}/k^{i,j}_{D,b} k_a=V^\star_{i,j}
k_a$ and $\tilde{k}_-^{i,j}\approx k_d$. Thus, in this case the rates
only depend on the microscopic reaction rates and of the
encounter volume $V^\star_{i,j}$.  In the right column of Fig.~9 (Fig.~9
b,d and f) a microscopic association rate of $k_a=8\text{ns}^{-1}$ is
chosen. This 8-fold increase in the microscopic association rate is
compensated by an 8-fold decrease in concentration. In this case the
reaction is strongly affected by diffusion ($k_a\geq k_{D,b}$) and the
macroscopic rates crucially depend on the different diffusive on- and
off- rates $k_D^{i,j}$ and $k_{D,b}^{i,j}$, which vary with the
diffusive properties of the different fragments as can be seen in
Tab. \ref{Tab.2} and a clearly different assembly dynamic is expected. We will
refer to this case as diffusion-influenced assembly.

In Fig.~9a and Fig.~9b the normalized fragment concentrations for the
assembly with reversible bonds ($k_d=0.0001\text{ns}^{-1}$) but
without an additional stabilization of the full ring structure
($k_a^\text{intra}=0$) is shown. Comparing the predictions from the
rate equation approach (Eq.~\ref{Eq. rate_eq}) (solid lines) with the
averaged simulation results (dashed lines) we observe excellent
agreement between our simulation results and the results from the
macroscopic rate equation approach in the reaction-limited regime as
well as in the strongly diffusion-influenced regime. The small deviations
between the simulation results and the rate equation approach in
Fig.~9a can be explained by our rough estimate for the excluded
volume. This agreement between the two approaches, especially for the
diffusion influenced regime is remarkable and demonstrates that, by
evaluating the relevant diffusive properties of the fragments, we are
indeed able to relate our microscopic reaction parameters to
macroscopic reaction rates that correctly reproduce the dynamics of
the system.  Comparing the equilibrium distribution of the cluster
fragments in Fig.~9a and Fig.~9b we see that indeed the 8-fold decrease
in concentration is balanced by the 8-fold increase in concentration,
as it was expected due to our previous reasoning. However, comparing
the kinetics of assembly we see a clear difference between Fig.~9a and
Fig.~9b. In general, the approach towards equilibrium is slower in
Fig.~9b than in Fig.~9a. This shows that we are indeed in a different
assembly regime and the change in concentration is not compensated by
the change in $k_a$ in the kinetics of the assembly process. At lower
concentration (Fig.~9b) clusters need a longer time to find each other
diffusively which is reflected in the slower approach towards
equilibrium.

In Fig.~9c and Fig.~9d the normalized fragment concentrations with an additional
stabilization of the final ring structure $k_a^\text{intra}=10
k_d=0.001$ns$^{-1}$ is shown. As discussed previously, the internal
bond formation can be understood as an intrinsic process which
stabilizes the full ring (Eq.~\ref{Eq. rate_eq2}). The value chosen for
the internal association rate corresponds to a free energy of
$E\approx - 2.3 k_B T$ associated with the bond formation according to
Eq.~\ref{Eq. internal}. In Fig.~9c and Fig.~9d, the normalized fragment
concentration of the full ring (black line) is shown irrespective of
the number of bonds in the ring. Comparing the simulation results and
the rate equation approach we again observe very good agreement
between them. Similarly to the case without internal bond formation
(Fig.~9a and Fig.~9b), the assembly dynamics at lower concentration
(Fig.~9d) is slower than at higher concentration (Fig.~9c), while the
equilibrium steady state remains the same in both cases. In contrast
to the reversible bond formation without additional stabilization of
the ring (Fig.~9c and Fig.~9e) the normalized concentration of the full
ring is significantly increased by the internal bond formation. Due to
mass conservation the increase in the concentration of the full ring
structure is balanced by a decrease in the concentration of smaller
fragments showing how an additional internal state can alter the
equilibrium cluster size distribution.

Finally, we investigate the assembly dynamics for irreversible bond
formation ($k_d=0\text{ns}^{-1}$) for two different microscopic
reaction rates $k_a$ corresponding to the regime of reaction-limited
reactions (Fig.~9e) and diffusion-influenced reactions
(Fig.~9f). Again the simulation results and the rate equation approach
agree remarkably well with each other. By comparing Fig.~9e and Fig.~9f
we again verify that the approach towards the steady state is much slower in the
case of lower concentration shown in Fig.~9f. 
However, in contrast to the previously discussed reversible
bond formation, the steady state is different in both cases with
different fraction of ring fragments containing 3,4 and 5
proteins. This reflects a fundamental difference between the steady
state observed for reversible bond formation and the steady state
observed for irreversible bond formation. In the case of reversible
bond formation the steady state is an equilibrium state in which
dissociation and association events balance each other. This state
only depends on the microscopic reaction rates $k_a$ and $k_D$ and the
size of the reactive volume $V^\star$ (Eq.~\ref{Eq. ka_prime}). In
contrast the steady state for irreversible bond formation is a trapped
steady state without any underlying dynamics. In this case of
irreversible bond formation the ring fragments cannot disassemble and
fragments of size $i\geq 5/2$ cannot further assemble if all smaller
fragments have been used up. Thus, in the case of trapping the final
state depends on the kinetics of the assembly. The difference in the
two steady states observed in Fig.~9e and Fig.~9f also shows that the
assembly dynamics in Fig.~9f is not only slowed down compared to
Fig.~9e, but that the different diffusive properties of the cluster
fragments lead to a change in the ratio of the concentration of
intermediates during the assembly process.  In general, larger
fragments are diffusing slower than smaller fragments, and hence the
rate for a reaction between two larger fragments is more strongly
affected by diffusion than the reaction rate for two smaller
fragments. In Fig.~9f this results in a lower concentration of
pentameric rings, as the reaction probability of two small fragments
is less affected by diffusion than the reaction probability of a
smaller and a larger fragment. This leads to a stronger depletion of
small fragments and hence a higher concentration of trapped ring
fragments with 3 or 4 proteins.

\section{Conclusion}

Biological structures like the actin cytoskeleton
\cite{Beltzner2008,Guo2009,Pollard2007} or the nuclear pore complex
\cite{Alber2007,Angelo2008,Tran2006} are highly dynamic with their constituents
being continuously exchanged. To understand the biological
functionality of these fascinating systems, a detailed understanding of
the assembly dynamics is crucial. Moreover, advances in the
fabrication of colloidal particles with directed interactions
(``patchy particles'') allow to design a plethora of differently
shaped building blocks with tunable interactions that can
self-assemble into new materials.
\cite{Mirkin1996,Michele2013a,Wang2012,Knorowski2011,Sacanna2010,Sacanna2013}
To increase the yield of the desired structures encoded in the local
particle interactions, kinetic trapping in unfavorable configurations
during the assembly process has to be prevented. Recently it has been
shown that particle interactions during the assembly process can be
actively controlled.  \cite{Leunissen2009,Michele2013} Thus, by a
state- or time-dependent switching of reactivity the assembly process
can be actively steered to reduce kinetic trapping. However, to fully
exploit the possibilities of controlling the assembly process, a
detailed understanding of its full dynamics with the formation of assembly
intermediates is necessary.

Here we presented a novel simulation approach which is ideally suited to
investigate the dynamics of large protein assemblies with well-defined
architectural properties. Proteins are represented by (multiple) non-overlapping,
hard spheres equipped with reactive patches. In contrast to most previous
studies on protein assembly
\cite{Rapaport2008,Hagan2006,Nguyen2007,Rapaport2012,Johnston2010,
Horejs2011,Wilber2007,Wilber2009,Liu2013,Sciortino2008,Michele2013}, which rely
on force fields to describe protein interactions, we
combine overdamped Brownian motion with the concept of reversible, stochastic
reactivity for patchy particles. Each cluster is treated as rigid object 
with its diffusive properties being evaluated on-the-fly. If the reactive patches of
two clusters form an encounter by force-free diffusive motion, a bond between
the clusters is established with a predefined rate. Local rules are used
to describe the resulting rearrangements, which are assumed to be
very fast on the time scale of the BD simulations. For potential-based
simulations, these rearrangements would proceed due to the 
corresponding forces, which do not exist in our approach. 
Similar to the association process, the dissociation process also
occurs with a predefined rate. Here we have derived the rules
for the placement of the two partners which are required to 
satisfy detailed balance. 

Together, these rules render our approach very efficient because 
complexes formed during the assembly are diffusing as
rigid objects (thus reducing the number of degrees of freedom to be propagated
to six) and because interactions are treated locally (as opposed to
evaluating potentials). Nevertheless, the patchy particle approach includes detailed information on the
architecture of protein assemblies, thus allowing us to study the
spatial-temporal dynamics of specific biological systems of interest.
Moreover, we have been able to show how the microscopic reaction
parameters used in our approach correspond to the macroscopic reaction rates
commonly used to describe the dynamics of the concentration of assembly intermediates.
This allows a direct comparison between macroscopic, experimentally measurable
concentrations and simulation results. To the best of our knowledge, this is the
first time that a BD algorithm combining stochastic reactivity of anisotropic
patches with reversible dynamics that fulfills detailed balance is presented.
Testing our algorithm against mean field prediction for the case of bimolecular
reactions revealed the importance of satisfying detailed balance.

As an example for a multi-component cluster, we investigated the
assembly of a pentameric ring. Here we compared our simulation results
to a rate equation approach for the different fragment
concentrations. By extracting the diffusive on- and off-rates from our
simulations we find excellent agreement between the rate equation
approach and our simulation results even for the case of
strongly diffusion-influenced assembly without any free parameter. We showed
that the assembly dynamics can be significantly changed by the change in the
diffusive properties of the assembly intermediates. In the case of
irreversible bond formation the system becomes trapped with
incompatible ring fragments. In this case not only the assembly
kinetics but also the finally reached steady state depends on the
different diffusive properties of the assembly intermediates. For
complex assembly geometries trapping is a common motif. Thus, in this
case not only correctly predicted equilibrium structure distribution
of the assembly intermediates is relevant but also the correctly
predicted kinetics of the assembly process as the system might become
trapped before reaching the expected equilibrium steady state. By
demonstrating how the macroscopic reaction rates can be calculated
from our simulation, we have developed a framework in which a
qualitative prediction of the changes in the macroscopic reaction
rates, based on the changes of the diffusive properties of the
assembly intermediates, is possible. 

Because our approach aims to describe the assembly of large
protein structures, the details of individual bond formations are not resolved within
this framework. In general, the use of local rules requires well-defined target
geometries and the instantaneous local rearrangement accompanying stochastic
bond formation follows from the assumption that the time scales of assembly and
local rearrangements are well separated. Given the coarse-grained nature of our
approach, the instantaneous rearrangement during binding  is a valid
approximation if the encounter volume specifies a narrow region around the
desired configuration. In the case of large encounter volumes the assumption of
fast and small local rearrangements can break down and the use of local rules
needs to be considered with care. In particular in the case of higher protein
densities or assembly close to a membrane this can lead to a situation in which
physically reasonable rearrangements are suppressed by our steric rules. In
classical MD or BD simulations with  potentials, these rearrangements would be
realized due to ensuing mechanical forces. Typical biological examples for
situations which are out of the scope of our current approach are the bending of
a membrane by BAR-proteins \cite{Arkhipov2008,Frost2009}, the emergence of
polymorphic virus structures \cite{Nguyen2008,Elrad2008,Nguyen2009} or the
assembly of actin networks from preexisting large fibers \cite{Blanchoin2014}. In principle, such a
situation could be resolved by using hybrid schemes that interface our approach
with potential-based simulations for local rearrangements. For the time being,
however, we focus on the assembly of protein complexes with well-defined
architectures, like native viruses\cite{Caspar1962}, centrioles
\cite{Kitagawa2011,vanBreugel2011,Gonczy2012,Guichard2013} or actin
filaments in networks growing by incorporation of actin monomers
from solution \cite{Guo2009,Blanchoin2014,Erlenkamper2013}.

In the future, our approach can be used to study the assembly of such
complex protein clusters. After we have successfully verified our
approach by comparing averaged simulation data with macroscopic mean
field results, we now can investigate the stochastic variance
inherent to these self-assembly processes. Our approach is
ideally suited to study the effect of time- or state-dependent changes
in reactivity, as has been recently shown in a qualitative study on
the effect on hierarchical assembly in virus
capsids. \cite{Baschek2012} Moreover, this approach might also help to design
artificial systems that do not get kinetically trapped in undesired
structures. It can also be coupled with hydrodynamic schemes, in
particular with reactive multi-particle collision dynamics \cite{Kapral2013}.

\section*{ACKNOWLEDGMENTS}

HCRK acknowledges financial support by the Cusanuswerk.
USS is a member of the Heidelberg cluster of excellence CellNetworks.
We thank Nils Becker and Joris Paijmans for helpful discussions.

\clearpage


%

\end{document}